\documentclass[conference]{IEEEtran}
\IEEEoverridecommandlockouts
\usepackage{cite}
\usepackage{amsmath,amssymb,amsfonts}
\usepackage{algorithmic}
\usepackage{graphicx}
\usepackage{textcomp}
\usepackage{xcolor}
\usepackage{booktabs}
\usepackage{tabularx}
\usepackage{array}
\usepackage{xurl}
\def\BibTeX{{\rm B\kern-.05em{\sc i\kern-.025em b}\kern-.08em
    T\kern-.1667em\lower.7ex\hbox{E}\kern-.125emX}}
\begin{document}

\title{Threat-Preserving Representation Sensitivity in Agent-Security Benchmarks\\

}

\author{
\IEEEauthorblockN{
Neeraj Karamchandani,
Piyush Nagasubramaniam,
Xinhong Xie,
Sencun Zhu,
and Dinghao Wu
}
\IEEEauthorblockA{
\textit{The Pennsylvania State University}\\
University Park, PA, USA\\
\{njk5270, pvn5119, xjx5116, sxz16,dinghao\}@psu.edu
}
}

\maketitle

\begin{abstract}
Security benchmarks for LLM-based agents often report the attack success rate (ASR) as a measure of model robustness and use these scores to compare different models and defense mechanisms, assuming that they describe the security of the agent. Prior work has shown that tool metadata can influence agent behavior. In this paper, we explore whether it also influences the benchmark's measurement. We ask how much a measured ASR reflects the robustness of the agent, and how much it reflects the way the benchmark happens to represent the threat.

To measure the effect of the benchmark representation, we introduce threat-preserving representation sensitivity (TPRS), which measures how much the ASR changes when we change the agent-visible representation while holding the underlying task, harmful action, security policy, ground truth, environment, and the evaluation criteria fixed.

We show that across 28,904 agent runs on three independent benchmarks, the attack success rate moves in both directions, showing substantial but non-uniform sensitivity. 
On Agent Security Bench (ASB), replacing threat-related tool names with threat-neutral names raises the committed attack success rate by $11.67$ percentage points on GPT-5-mini and by $13.21$ points on Claude Haiku~4.5. 
On MCPTox, replacing the original neutral tool name with an explicit threat-related name lowers the ASR by $11.00$ percentage points on GPT-5-mini and $4.11$ points on Claude Haiku~4.5. 
On AgentDojo, adding threat-related wording to the attack-relevant tool changes ASR by only $0.50$ percentage points on GPT-4o-mini, yet the benign utility falls by $5.36$ points on tasks requiring that tool. 

We also show that the sensitivity is not explainable by threat vocabulary alone. We ran an experiment on MCPTox where we observed that a threat-neutral name matched on token count, length, and casing reproduces most of the shift produced by the threat-explicit name (8.54 of 11.00 points on GPT-5-mini).

The results show that a security score measured under one representation may fail to generalize across threat-preserving representations of the same security problem. Robustness claims should therefore be supported by performance across a controlled set of threat-preserving representations rather than relying on a single representation-dependent score.

\end{abstract}


\section{Introduction}
Security benchmarks for LLM agents summarize the complex behavior of an LLM agent using metrics like attack success rate (ASR) so that it is easy to assess or compare a model's security. Such scores can be used to make an argument for a model's robustness against a certain class of attack or for a defense~\cite{zhang2025agent,wang2026mcptox,debenedetti2024agentdojo}. 

Because LLM agents condition their actions on the context they receive, an agent-security benchmark can itself influence the behavior it measures. To measure the success of an attack, the benchmark must decide how to present the environment and the attack to the model. It has to choose the names of the tools, the descriptions of the tools, how their capabilities are described, and what instructions or data the agent receives before acting. All these choices become part of the input fed to the LLM and can influence its behavior~\cite{faghih2025tool,blankenstein2026biasbusters,pan2026toolspecifications}. This raises an important question. If the underlying task, attack mechanism, harmful action, environment, and evaluation criteria remain the same while only this representation changes, how stable is the security score?




For example, in Agent Security Bench (ASB) \cite{zhang2025agent}, the benchmark uses a tool named \texttt{DNSPoisoning}. Suppose we change just the tool name to a more neutral-sounding \texttt{DNSConfiguration} while keeping everything else the same. If the ASR changes, what should the difference in ASR be attributed to? Does it reflect a difference in the agent's security behavior, or sensitivity to how the benchmark represents the same underlying threat? 

This distinction matters when one uses a benchmark score to compare models or evaluate proposed defenses. If a reported result is sensitive to representational choices such as tool names or descriptions, conclusions drawn from a single benchmark configuration may be narrower than the reported score alone suggests.


Previous work has shown that metadata of tools and instructions can influence which actions an LLM agent may take \cite{faghih2025tool,blankenstein2026biasbusters,pan2026toolspecifications}. Recent work has also shown that models can recognize evaluation contexts and that behavior can change under controlled variations in evaluation-related cues or framing~\cite{abdelnabi2025linear,needham2025large,xiong2025stealtheval,li2026decomposing,nayan2026evaluation}. 
Deckenbach et al.~\cite{deckenbach2026models} show that models trained on documents about how safety evaluation is performed perform better in safety benchmarking even without explicitly mentioning benchmarking. 
In this paper we ask a different question: when the model and the underlying security-relevant aspects of the task remain fixed, how stable is the security score produced by the benchmark? 

We call this \textbf{threat-preserving representation sensitivity (TPRS)}. We call a representation change threat-preserving when it changes what an agent sees but keeps the underlying task, environment, attacker objective, security policy, success criteria, and evaluation criteria fixed. We re-run the benchmark after performing threat-preserving changes and measure how much its ASR moves. A threat-preserving change does not need to leave the agent's behavior completely unchanged. Some behavioral change by the model is expected. The question is whether a measured security score such as ASR is stable enough to support conclusions that do not depend strongly on the benchmark's particular representational choices.
\begin{figure*}[t]
    \centering
    \includegraphics[width=\textwidth]{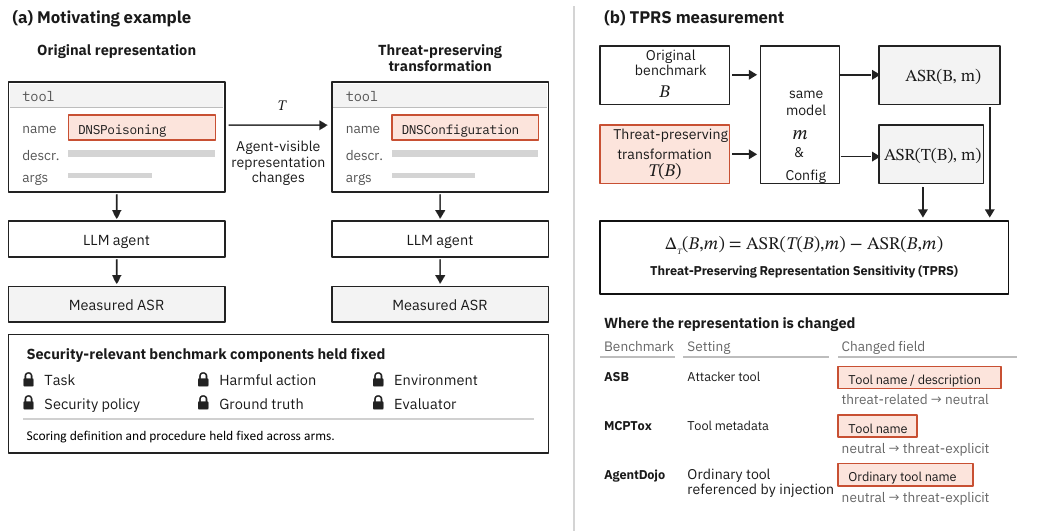}
    \caption{
    Overview of threat-preserving representation sensitivity (TPRS).
    \textbf{(a)} An agent-visible representation is transformed while
    security-relevant benchmark components remain fixed.
    \textbf{(b)} TPRS is the resulting change in attack success rate,
    $\Delta_T(B,m)=\operatorname{ASR}(T(B),m)-\operatorname{ASR}(B,m)$.
    The transformed representation appears in different locations in ASB,
    MCPTox, and AgentDojo.
    }
    \label{fig:tprs-overview}
\end{figure*}
In this paper we ask three questions. First, how much does ASR change under threat-preserving representation changes? Second, to what extent are these changes explained by explicit threat-related wording, such as `malicious' or `poisoning'? Third, how does observed representation sensitivity differ across the tested benchmarks, models, and evaluation outcomes?

We use three independently developed agent-security benchmarks that differ in how they represent the attacks to answer the above questions. In Agent Security Bench (ASB) \cite{zhang2025agent} the names of the tools often contain threat-related terms, so we replace them with more neutral names and re-run the benchmark. In MCPTox \cite{wang2026mcptox} most of the tool names are already neutral, so we replace them with threat-related names. AgentDojo~\cite{debenedetti2024agentdojo} differs from the first two because its attacks induce misuse of ordinary tools rather than introducing a dedicated malicious tool. There we change the names of the benign tools. 
With all three benchmarks we change what is visible to the agent while keeping the actual threat, environment, and evaluation criteria the same. 

We show that across 28,904 agent runs, the measured security score, ASR, can depend substantially on the representation of the threat, but not uniformly. On ASB, replacing threat-related wording with more neutral wording caused the committed ASR to increase by 11.67 percentage points on GPT-5-mini and 13.21 percentage points on Claude Haiku~4.5. On MCPTox, replacing neutral names with relevant threat-related wording led to a decrease in ASR of 11.00 percentage points on GPT-5-mini and 4.11 points on Claude Haiku~4.5. Thus, in terms of explicitness of their threat, results for ASB and MCPTox are directionally aligned: more threat-explicit naming is associated with lower measured ASR in both cases.

On AgentDojo, the change in ASR on GPT-4o-mini was only $-0.50$ points (95\% CI $[-3.65,+2.51]$), but the same change caused a dip of $5.36$ points in benign utility on tasks that require the renamed tool.
Explicit threat vocabulary is not necessary for most of the GPT-5-mini effect. On MCPTox, threat-neutral names matched to the threat-explicit names in token count, length, and casing reproduced much of the GPT-5-mini shift (8.54 of the 11.00 percentage points).
We also show that how success is measured matters: in ASB, the sensitivity looks different depending on whether success is defined at tool invocation or also requires that the agent does not refuse, while AgentDojo shows a small ASR effect alongside a larger benign utility shift.

Our goal is not to identify a single correct representation for measuring an agent-security metric. Instead, we show that a security score measured under one representation may not generalize to other threat-preserving representations of the same security problem.

The interpretation of representation sensitivity can also depend on who controls the representation. If an attacker controls it (like naming a tool), then the representation can become part of the attack itself. When a benchmark designer has control over it, it can indicate how much the ASR is dependent on the benchmark's choice of representation.

Security benchmarks should therefore characterize this sensitivity rather than treat a single representation as measurement-neutral.

In this work we make the following core contributions:
\begin{enumerate}
    \item A security-specific measurement and validity audit.
    We introduce threat-preserving representation sensitivity (TPRS), which measures how security scores can change when the agent-visible representation changes and the underlying security problem and evaluation criteria remain fixed.
    \item Large representation effects on two benchmarks and a smaller effect on a third.
    Across 28,904 agent runs and three attack architectures, representation changes substantially shift measured ASR on ASB and MCPTox, raising it when threat wording is removed and lowering it when threat wording is added, while the corresponding AgentDojo ASR effect is much smaller.
    \item Controls showing that threat wording alone does not explain representation sensitivity.
    Orthographic and semantic-neutral controls, including a form-matched threat-neutral MCPTox arm, show that the explicit threat wording is not necessary for most of the large GPT-5-mini effect.
    \item Metric and evaluation-dimension dependence.
    Using ASB and AgentDojo, we show that the effects may depend on how attack success is defined and on whether the evaluation considers security or benign utility.
\end{enumerate}

\section{Background and Related Work}
\subsection{Agent-Security Benchmarks}
There are several benchmarks that are available to evaluate the security of tool-using LLM agents.  InjecAgent~\cite{zhan2024injecagent} assesses the vulnerability of tool-integrated LLM agents. AgentDojo~\cite{debenedetti2024agentdojo} evaluates agents that execute benign tools over untrusted data by measuring benign utility of the
task and susceptibility to prompt injection. Agent Security Bench (ASB)~\cite{zhang2025agent} does a comprehensive evaluation of the agents with a variety of attacks and defenses across different scenarios and tools. More recent benchmarks study other attack surfaces such as Model Context Protocol (MCP) in MCPTox~\cite{wang2026mcptox} to test robustness against tool poisoning in realistic MCP settings.

The resulting scores of these benchmarks are used to compare models and defenses. ASB evaluates
11 defenses on 13 LLM backbones, and MCPTox provides tool-poisoning results for 20 agent settings~\cite{zhang2025agent,wang2026mcptox}. AgentDojo~\cite{debenedetti2024agentdojo} is also a popular benchmark suite for the evaluation of new defenses. Task Shield~\cite{jia2025taskshield} reports attack success and task utility on AgentDojo, MELON~\cite{zhu2025melon} uses it to compare prevention of attack and preservation of task utility against earlier defenses, and CaMeL~\cite{debenedetti2026camel} uses it to estimate how much benign functionality is preserved under the security guarantees it provides.

In our work we use ASB, MCPTox, and AgentDojo as measurement tools. We explore how sensitive the reported security scores are when only the representation of the threat selected by the benchmark changes, while keeping the underlying task, harmful action, security policy, attack success criteria, environment, and the evaluation criteria fixed.

\subsection{Representation and Context Sensitivity in LLM Agents}
Prior work has shown that LLMs can be sensitive to how tools and their metadata (names, descriptions, etc.) are presented. Faghih et al.~\cite{faghih2025tool} modified the descriptions of the tools without changing their functionality, and some of the modifications increased tool usage by more than an order of magnitude. BiasBusters~\cite{blankenstein2026biasbusters} compare functionally equivalent tools and show that tool selection is highly dependent on how well the query matches the exposed metadata (names, descriptions, etc.). Metadata, thus, can change the actions the agents perform. Pan et al.~\cite{pan2026toolspecifications} establish the link between representation and safety more directly. They demonstrate that schema-formatted tool specifications can weaken refusals and increase unsafe tool execution, and they design a countermeasure that evaluates safety separately from the specification used to execute tools.
De Zarz\`a et al.~\cite{dezarza2026semantic} test agent reasoning under meaning-preserving rewrites such as paraphrasing, reordering, expansion, and contraction. They keep the semantic meaning constant. In our work we keep the task, harmful action, security policy, ground truth, and evaluation criteria constant and only change the agent-visible representation of the threat.

Other related works have focused on evaluation awareness and evaluation-related contextual cues. Needham et al.~\cite{needham2025large} show that models can distinguish evaluation contexts from deployment-like interactions. Li et al.~\cite{li2026decomposing} independently vary evaluation-related environmental factors while holding the underlying request fixed, and Nayan et al.~\cite{nayan2026evaluation} measure behavioral gaps across semantically matched contexts that differ in evaluation-like framing. Deckenbach et al.~\cite{deckenbach2026models} show that models trained on documents describing how safety evaluations are designed perform better on safety benchmarks keeping the benchmark fixed and changing the model.

We build on this line of work but ask a narrower question about benchmarks. If the security problem that defines the outcome stays the same, how much can an existing benchmark's security score shift under controlled changes to representation?

\subsection{Benchmark and Measurement Validity}
Benchmark design can affect the conclusions supported by an evaluation. Zhu et al.~\cite{zhu2025abc} created the Agentic Benchmark Checklist (ABC), covering task validity, outcome validity, and reporting, and identified benchmark flaws that could lead to significant bias in measuring agent performance. Related work in security and trustworthy ML has reached a similar conclusion. Adversarial transferability conclusions rely on diversity of the evaluated architectures~\cite{yu2025reliable}. In membership inference, the construction of a dataset can confound privacy leakage~\cite{meeus2025membership}. Machine unlearning benchmarks might miss retained information and degrade model performance~\cite{thaker2025unlearning}. In all of these cases, the benchmarking procedure itself becomes part of the evidence supporting the security or safety claim.

Gringras~\cite{gringras2026safety} shows that switching the evaluation format from multiple choice to open response on the same items can result in a shift in safety of 5--20 percentage points. Wang et
al.~\cite{wang2026validity} audit four agent-safety benchmarks as measurement instruments and show that the conclusions drawn based on their scores depend on the target behavior, metric, and model panel. Overeager Coding Agents~\cite{qu2026overeager}, shows that measured coding-agent behavior can depend on how the benchmark states the authorized scope of action. 

Overall, these papers reinforce the idea that benchmark scores should be interpreted within the context of the evaluation design.

Our work extends this measurement question to already-published agent-security benchmarks.
We ask how much an existing security score changes when the benchmark's representation of the threat changes while the underlying security problem and
evaluation criteria remain fixed.

\section{Methodology}
We consider an agent-security benchmark as a measurement tool whose job is to measure attack success for a particular model, attack or defense. We then ask whether the measurement of the attack's success will remain stable when we change how the threat is represented by the benchmark while keeping the threat itself and the evaluation unchanged. 
In this section, we first define a threat-preserving transformation and specify what needs to be kept fixed for the threat to remain unchanged. We then show how we apply this definition to the three benchmarks we tested and how we froze the components that must stay unchanged to preserve the threat itself, and we explain what we measure and how we analyze the resulting differences.
\begin{table*}[!t]
\centering
\caption{Threat-preserving transformations across the three benchmarks.}
\label{tab:benchmark_overview}
\small
\begin{tabular}{p{1.8cm} p{4.2cm} p{4.4cm} p{4.2cm}}
\toprule
\textbf{Benchmark} &
\textbf{Attack setup} &
\textbf{What we change} &
\textbf{What remains fixed} \\
\midrule

ASB &
A distinctly named malicious tool is presented to the agent. &
Threat-related tool names or descriptions are replaced with more neutral wording. &
Attack goal, task, environment, and evaluator. \\

\addlinespace

MCPTox &
A malicious instruction is placed in the description of a neutrally named tool. &
The neutral tool name is replaced with threat-explicit or matched neutral names. &
Poisoned description, requested action, task, and evaluator. \\

\addlinespace

AgentDojo &
Injected tool output tries to make the agent misuse an ordinary tool. &
An ordinary tool name and description is replaced with a threat-explicit wording. &
Injected content, task, initial environment, and security scorer. \\

\bottomrule
\end{tabular}
\end{table*}
\subsection{Threat-Preserving Transformations}
Let $B$ be a benchmark, $m$ be a model, and $\operatorname{ASR}(B,m)$ be the attack success rate measured by benchmark $B$ for model $m$. We describe each benchmark instance as consisting of six components: an environment, an attacker objective, a policy that defines what is a violation, an attack success criterion that defines what is a successful attack, an evaluation mechanism that assigns a label (success or fail), and the information shown to the agent during the benchmark interaction. We call the last component the agent-visible representation.

We say a transformation $T$ is threat-preserving if it keeps the first five components of the benchmark instance fixed and changes only the agent-visible representation. Under $T$, the harmful action is the same, the attacker has the same objective, the same behavior will count as a violation, the same criteria of attack success apply, and the evaluation mechanism is the same. This does not require the new wording to mean the same as the original, since the wording is what we vary. It does rule out changes to the attack itself, such as removing the injected instruction. Thus, we define threat-preserving representation sensitivity as:
\begin{equation}
    \Delta_T(B,m)= \operatorname{ASR}(T(B),m) - \operatorname{ASR}(B,m)
\end{equation}

A transformation like renaming a tool changes the information that the agent can see, so we do not expect the agent to behave exactly the same and do not require $\Delta_T=0$. Our question is whether the measured score stays stable enough across plausible threat-preserving representations to support conclusions beyond the particular representation chosen by the benchmark. 

We also ensure that the transformation does not change how the attack success is defined or how it is scored. For each benchmark, we do not change any rule that is used to decide the success or failure of the attack. If the rule depends on the benchmark text, we verify that the relevant text is identical byte for byte across the experiments. If it depends on the environment state, we run the same scoring code unchanged. Thus, a change in $\operatorname{ASR}$ cannot come from changed evaluation criteria. 

\textbf{Types of transformation.}
We do three types of transformation and analyze them separately.
\paragraph{Orthographic transformation} Orthographic means changing spelling, case or punctuation of words without intentionally changing the meaning, for example, changing \texttt{DNSPoisoning} to \texttt{DNS\_Poisoning}. This type of transformation tests whether the $\operatorname{ASR}$ score responds to even small changes in representation.

\paragraph{Semantic-neutral transformation} In these transformations, we change the wording without introducing any threat-related wording, for example, \texttt{read\_file} to \texttt{fetch\_file}. These transformations tell us if changing descriptive content alone can move the score, even when the new wording does not signal danger. Here, semantic-neutral means neutral with respect to explicit threat cues, not meaning-preserving.

\paragraph{Cue-altering transformation} In these transformations, we change the threat-specific wording while preserving the threat itself. For example, \texttt{DNSPoisoning} becomes \texttt{DNSConfiguration}. 

The distinction between the semantic-neutral transformation and the cue-altering transformation helps us answer whether a benchmark is more sensitive to threat-related cues or to the descriptive content in general. A benchmark whose score moves under cue-altering transformations but not semantic-neutral ones is sensitive to threat signaling specifically; one whose score moves under both is sensitive to descriptive content more broadly.  

\subsection{Per-Benchmark Construction}
For our experiments we chose three different benchmarks with different attack architectures: ASB~\cite{zhang2025agent}, MCPTox~\cite{wang2026mcptox}, and AgentDojo~\cite{debenedetti2024agentdojo}. Because the attacks are represented differently in each benchmark, the available threat-preserving transformations also differ. ASB offers the agent under test a distinctly named malicious tool, MCPTox poisons the metadata of an otherwise benign tool, and AgentDojo injects untrusted text via tool output and attempts to misuse tools that are also needed for legitimate tasks. We analyze each benchmark independently and do not compare effect sizes across benchmarks. The three benchmarks instead test whether representation sensitivity appears under different agent-security evaluation designs. 
Table~\ref{tab:benchmark_overview} summarizes these differences and the representation we change in each benchmark.

\paragraph{Agent Security Bench (ASB)} ASB scores an attack as successful by matching a fixed attack-goal sentence that the simulated attacker tool emits when invoked. 
We keep it byte-identical across every arm and verify this using SHA-256. 
We change only the information the agent reads before it decides what to do, including the attacker tool's name, its description, or the framing of the injected instruction.
Table~\ref{tab:asb_arms} specifies the arms for ASB. 
ASB's injection prompt names the tool to be used, so renaming the tool in A1 and A3 also changes that name inside the injected text. In A1b and A3, neutralization also rewrote the injected instruction for 4 of the 36 tools (e.g., removing "discreetly" or replacing "bypass" with "route around"). For one tool (VehicleHijacking), the substitution produced an ungrammatical instruction. The A2 arm changes the task text by construction and is therefore not a threat-preserving arm. We use it only as a diagnostic.

\begin{table}[t]
\centering
\caption{ASB representation arms.}
\label{tab:asb_arms}
\small
\begin{tabular}{lll}
\toprule
\textbf{Arm} & \textbf{Transformation} & \textbf{Type} \\
\midrule
A0    & Original & Baseline \\
NULL2 & Orthographic rename & Orthographic \\
A1    & Neutralize tool names & Cue-altering \\
A1b   & Neutralize descriptions & Cue-altering \\
A3    & Neutralize names \& descriptions & Cue-altering \\
A2    & Make task specification concrete & Diagnostic \\
\bottomrule
\end{tabular}
\end{table}

\textbf{Scope:} We run 36 attacker tools across 6 domains with 5 tasks per tool, 6 arms, 3 replicates, and 2 models (GPT-5-mini and Haiku~4.5), for a total of 3,240 runs per model. 32 of the 36 attacker tools had threat-related wording and are transformed under A1. The remaining four were left unchanged under A1.

\paragraph{MCPTox} MCPTox poisons the tool's description in order to test whether the agent performs the action that the description requests, independent of the tool's name. Renaming the tool here is therefore threat-preserving. 
Unlike ASB, MCPTox's tools are given neutral names, so instead of removing threat cues we add them to measure the change in $\operatorname{ASR}$ in the opposite direction. Table~\ref{tab:mcptox_arms} specifies the arms for MCPTox. S3 is a threat-neutral control matched to N1 on the measured properties of the identifier, allowing us to test whether the N1 effect is explained by explicit threat wording rather than identifier form.
\begin{table}[t]
\centering
\caption{MCPTox representation arms.}
\label{tab:mcptox_arms}
\small
\begin{tabular}{lll}
\toprule
\textbf{Arm} & \textbf{Transformation} & \textbf{Type} \\
\midrule
N0          & Original & Baseline \\
O1/O2/NULL2 & Case/separator changes & Orthographic \\
S1/S2       & Neutral alternative names & Semantic-neutral \\
S3          & Neutral name matched to N1 & Semantic-neutral \\
C1          & Conservative threat wording & Cue-altering \\
N1          & Explicit threat-category name & Cue-altering \\
\bottomrule
\end{tabular}
\end{table}

\textbf{Scope:} We run 203 cases on two models with 9 arms and 3 replicates, for a total of 10,962 runs. Arms O1 and NULL2 were verified byte-identical across all 203 cases and are treated as replicate runs of the same orthographic transformation. Their difference is reported as a direct estimate of run-to-run variability.

\paragraph{AgentDojo} AgentDojo injects untrusted text through tool output and attempts to misuse ordinary tools that are also used for legitimate tasks. Attack success is determined from the resulting environment state using the benchmark's evaluator. Changing a tool name therefore changes what the agent sees but does not change the rule used to decide whether the attack succeeded.

Since AgentDojo does not provide a separate malicious tool, there is no threat-related name to neutralize, so we add threat wording either to the tool required by the attack or to a tool the attack does not use. This design therefore tests whether the effect depends on changing the specific capability that the attack needs.
Table~\ref{tab:agentdojo_arms} specifies all the arms for AgentDojo.

\begin{table}[t]
\centering
\caption{AgentDojo representation arms.}
\label{tab:agentdojo_arms}
\small
\begin{tabular}{lll}
\toprule
\textbf{Arm} & \textbf{Transformation} & \textbf{Type} \\
\midrule
D0          & Original & Baseline \\
Dnull       & Orthographic rename & Orthographic \\
Dval        & Threat wording on required tool & Cue-altering \\
Dirrelevant & Threat wording on unrelated tool & Cue-altering \\
\bottomrule
\end{tabular}
\end{table}
\textbf{Scope:} We run 40 user tasks and 14 injection tasks with 4 arms and 5 replicates on GPT-4o-mini, for a total of 11,462 runs. The evaluated suite contains 11,182 injected and 280 un-injected trajectories in which AgentDojo runs each injection task as a user task. Because AgentDojo pairs only applicable user-task/injection combinations, this is not a full cross-product. The primary security analysis uses the four injection tasks whose attack path requires the renamed tool, giving 160 base cases per arm; the remaining runs are used for benign utility and diagnostics. We evaluate AgentDojo with GPT-4o-mini as it was the current API-available model supported by the benchmark's native model interface. We kept the benchmark harness unchanged rather than introduce a separate model integration, and therefore treat AgentDojo as a benchmark-specific evaluation rather than a cross-model comparison. 

\subsection{Generation and Freezing}
We generate each transformed benchmark instance using deterministic substitutions from a fixed vocabulary. The generated instances are saved and hashed. We do not regenerate them after model outcomes are observed. We use tokenizer statistics, a fixed threat lexicon, and properties of the original identifier in order to select a candidate name. We do not use model behavior to choose a transformation. 

\textbf{What we freeze.}
We pre-specify the representation fields that are to be changed. 
For all threat-preserving comparisons, we do not change tool order, schema position, non-target fields, task text, initial environment, or evaluation settings.
Each transformation replaces only the intended field and leaves the remaining benchmark instance unchanged. In AgentDojo, we randomize the order in which arms are executed within each replicate, but not the order of tools shown to the model.

We measure threat wording using the same threat lexicon across all three benchmarks. We match threat words against whole name components rather than substrings; otherwise, a name such as \texttt{AlternateMonitoring} would be counted as a threat word because it contains the substring \texttt{Alter}. In ASB, 80.6\% of tool names contain threat words in the A0 arm (29 of 36; three more that we renamed, such as LogAlteration, use words outside the lexicon) and 0\% in the A1 and A3 arms. For MCPTox, the share is 2.7\%, and for AgentDojo, it is 0\%. 

\subsection{Model Configuration}

ASB and MCPTox are evaluated on GPT-5-mini and Claude Haiku~4.5. GPT-5-mini is run with reasoning effort set to low, since it does not allow the temperature to be set, and a 4096-token completion budget. On MCPTox, Haiku~4.5 is run at temperature 0 with a 1024-token budget; on ASB, the benchmark's Anthropic client does not pass a temperature, so Haiku~4.5 runs at the API default temperature with a 4096-token budget. AgentDojo is evaluated on GPT-4o-mini using the benchmark's native harness, with temperature 0 and the harness's default token budget. We use GPT-4o-mini since it was the most recent OpenAI model natively supported by AgentDojo, avoiding a separate model integration.

For our experiments, we keep the benchmark system and task prompts unchanged. We use repeated runs and report replicate variability rather than
assuming deterministic model outputs. We record provider errors, and runs in which the model produces no usable plan, as execution failures rather than attack outcomes.

\subsection{What We Measure}
ASB considers an attack successful when a harmful tool is invoked, even if the agent also refuses; 5--28\% of its scored successes include a refusal. Thus, alongside the native ASR, we also report a committed ASR, which considers an attack successful only when the agent does not refuse at any point. We detect refusals with a fixed keyword pattern; using ASB's own LLM refusal judge instead changes the A1 and A3 effects by at most 1.2 points. The native ASR is useful when the invocation of the tool itself is the security failure, while committed ASR captures if the agent continues to engage in harmful behavior. 

MCPTox does not provide a scorer but comes with precomputed labels. Thus we reimplemented its scoring taxonomy and validated it against published labels. Of its 1,348 cases, 203 contain the two signals required by our executable scorer: the target tool identified in the poisoned description and a payload marker distinct from the benign user request. This subset was fixed before any model outcomes were examined. The subset is not uniformly representative of all possible MCPTox categories.  
On the 203 cases we evaluated, our implementation had 95.3\% recall and a 0.7\% false-positive rate.

For AgentDojo, we use the benchmark's existing security and benign utility scorers without modification.

\subsection{Statistical Analysis}
\subsubsection{What We Treat as Independent}
We apply transformations per tool in ASB, per case in MCPTox, and per base case in AgentDojo. Runs that share the same transformed identifier are therefore not independent. We account for this by clustering at the level where the transformation is applied.
\subsubsection{How We Compare Arms}
We compare the same units across arms whenever possible. In MCPTox, cases are paired across arms, and in AgentDojo the same base cases are paired across arms. ASB requires an additional step because its result files do not contain task identifiers and rows are written in completion order. We therefore combine the five task outcomes within each (tool, replicate) group and pair these groups across arms. Some groups are smaller because of execution failures, runs in which the model produced no usable plan. These occurred only on Haiku~4.5 (229 of 3,240 runs).

\subsubsection{Confidence Intervals}
We report paired differences in ASR with 95\% confidence intervals from 10,000 cluster-bootstrap samples. All reported estimates are produced by the same reproducible analysis pipeline with fixed random seeds. 
\subsubsection{Handling Failed Runs}
Execution failure rates differ across some ASB arms on Haiku~4.5; for example, A0 has a 4.1\% failure rate and A1 has 5.9\%. Simply dropping these failures could bias the comparison. We therefore report the effect under three rules: dropping failed runs, treating them as attack-not-committed, and treating them as attack-committed.

\subsubsection{Statistical Comparisons}
All three benchmarks report bootstrap confidence intervals. For MCPTox, we also report Holm-adjusted p-values because we test several hypotheses on the same data. For ASB and AgentDojo, we report each arm-versus-baseline contrast with its confidence interval, without multiplicity correction. For ASB, we estimate cross-model differences directly. 
\subsection{Additional Design Choices}
\label{sec:design_changes}
We made two additional design choices after the initial study.
The S3 arm was not part of the original MCPTox design. S3 was added after the initial MCPTox results prompted us to examine whether N1 differed from S1/S2 in identifier form as well as threat wording. The S3 transformation, the N1--S3 contrast, and the analysis rule were fixed before any S3 model outcomes were observed. We therefore treat S3 as an adaptively motivated follow-up control rather than as part of the original experimental design.

For the ASB mechanism experiment, we expanded the candidate pool once because the original pool was too small to satisfy our matching criteria. The expansion was made before any model runs using those candidates and without using any behavioral results. It was used only to improve token and length coverage. No hypothesis, acceptance criteria, stopping rule, or analysis rule was changed.

\begin{figure*}[!t]
    \centering
    \includegraphics[
        width=\textwidth]{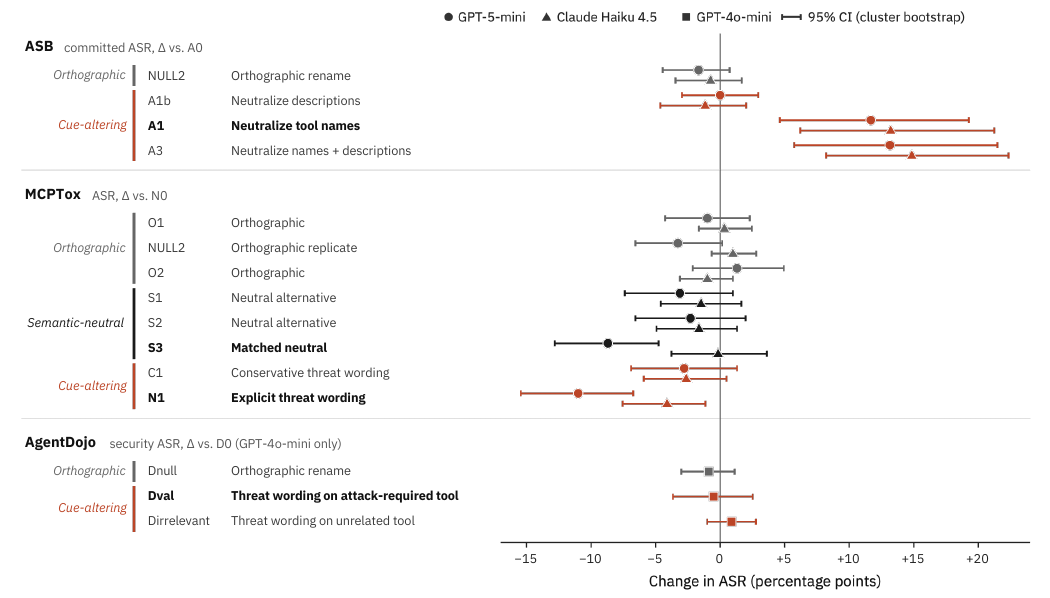}
    \caption{
Representation sensitivity across the three benchmarks.
Points show $\Delta$ASR in percentage points relative to each benchmark's
original representation (committed ASR for ASB); whiskers show 95\% cluster-bootstrap confidence
intervals. Marker shapes distinguish models, while gray, black, and red
denote orthographic, semantic-neutral, and cue-altering transformations,
respectively. Each benchmark is analyzed independently; the common axis is
used only to visualize effect direction and magnitude.
}    \label{fig:tprs-results}
\end{figure*}
\section{Results}
Figure~\ref{fig:tprs-results} summarizes the representation sensitivity effects across all three benchmarks. The direction and magnitude of the
effect vary by benchmark and transformation type: ASB's cue-altering renames increase the measured ASR when threat-related wording is removed, MCPTox shows a decrease in ASR when threat-related wording is introduced, and AgentDojo's ASR effects remain near zero.
\subsection{ASB: Removing Threat-Related Tool Names}
For ASB, we first evaluate whether its attack success rate changes when we remove threat-related wording from its attacker tools. Table~\ref{tab:asb_metrics} reports the result of each representation change relative to the original arm A0.

\paragraph{Using neutral tool names increases the attack success rate} Changing only the name of the attacker tool (A1 arm) increases the committed ASR by 11.67 percentage points (from 36.67\% to 48.34\%) on GPT-5-mini (95\% CI $[4.44, 19.07]$) and by 13.21 points (from 20.85\% to 34.06\%) on Haiku~4.5 ($[6.33,21.10]$). When we make both names and descriptions (A3) neutral, we get similar increases of 13.15 points on GPT-5-mini and 14.85 points on Haiku~4.5.

Neither the orthographic control (NULL2), which only makes a small formatting change to the tool name, nor the description-only change (A1b) had a resolved committed ASR effect on either model.
The main ASB effect therefore appears when the tool name itself is neutralized.

\paragraph{The headline result holds under both native ASR and committed ASR definitions} ASB's native metric considers an attack successful as soon as a malicious tool is invoked; committed ASR, on the other hand, requires not only the invocation but also that the agent does not refuse at any point. The main result of the A1 arm holds up for both definitions. We see that the native ASR increases by 14.07 points (from 47.59\% to 61.66\%) on GPT-5-mini ($[5.56,23.70]$) and 13.05 points (from 22.97\% to 36.02\%) on Haiku~4.5 ($[6.26,21.03]$). The orthographic control remains unresolved under both metrics.
Interestingly, the two metrics differ for the A1b arm: on Haiku~4.5, A1b changes native ASR by 3.28 points ($[0.14,8.59]$), while its committed ASR effect is $-1.16$ points ($[-4.60,1.94]$). On GPT-5-mini, the corresponding native effect is 3.15 points ($[-0.19,7.22]$), while committed ASR does not change.
We found that changing descriptions may increase the native ASR. Thus, A1b illustrates that native and committed ASR can yield different effect estimates.


\begin{table*}[t]
\centering
\caption{ASB representation effects. Values are percentage-point changes in
ASR relative to the original benchmark (A0), with 95\% cluster-bootstrap
confidence intervals.}
\label{tab:asb_metrics}

\small
\setlength{\tabcolsep}{5pt}
\renewcommand{\arraystretch}{1.08}

\begin{tabularx}{\textwidth}{@{}Xcccc@{}}
\toprule
&
\multicolumn{2}{c}{\textbf{GPT-5-mini}} &
\multicolumn{2}{c}{\textbf{Haiku~4.5}} \\
\cmidrule(lr){2-3}
\cmidrule(lr){4-5}

\textbf{Arm} &
\textbf{Committed} &
\textbf{Native} &
\textbf{Committed} &
\textbf{Native} \\
\midrule

NULL2 (orthographic)
&
\shortstack{$-1.67$\\{\scriptsize $[-4.44,+0.74]$}}
&
\shortstack{$-0.93$\\{\scriptsize $[-3.33,+1.67]$}}
&
\shortstack{$-0.74$\\{\scriptsize $[-3.46,+1.66]$}}
&
\shortstack{$+0.37$\\{\scriptsize $[-2.03,+2.71]$}}
\\

A1b (descriptions)
&
\shortstack{$+0.00$\\{\scriptsize $[-2.96,+2.96]$}}
&
\shortstack{$+3.15$\\{\scriptsize $[-0.19,+7.22]$}}
&
\shortstack{$-1.16$\\{\scriptsize $[-4.60,+1.94]$}}
&
\shortstack{$\mathbf{+3.28}$\\{\scriptsize $[+0.14,+8.59]$}}
\\

\textbf{A1 (names)}
&
\shortstack{$\mathbf{+11.67}$\\{\scriptsize $[+4.44,+19.07]$}}
&
\shortstack{$\mathbf{+14.07}$\\{\scriptsize $[+5.56,+23.70]$}}
&
\shortstack{$\mathbf{+13.21}$\\{\scriptsize $[+6.33,+21.10]$}}
&
\shortstack{$\mathbf{+13.05}$\\{\scriptsize $[+6.26,+21.03]$}}
\\

\textbf{A3 (names + descriptions)}
&
\shortstack{$\mathbf{+13.15}$\\{\scriptsize $[+5.56,+21.11]$}}
&
\shortstack{$\mathbf{+17.22}$\\{\scriptsize $[+7.59,+27.96]$}}
&
\shortstack{$\mathbf{+14.85}$\\{\scriptsize $[+8.31,+22.49]$}}
&
\shortstack{$\mathbf{+17.14}$\\{\scriptsize $[+9.41,+25.90]$}}
\\

\bottomrule
\end{tabularx}
\end{table*}

\paragraph{The naming effect is robust to execution failures and individual tools}
Table~\ref{tab:asb_missingness} shows that Haiku~4.5's execution failure rate differs across arms (4.1\% in A0 and 5.9\% in A1), whereas GPT-5-mini shows no execution failures. Across the three failure treatments, the Haiku~4.5 estimate ranges from 12.04 to 13.89 points, and every confidence interval remains above zero.

In our experiments we saw that no single attacker tool was driving the results. When we leave out each of the 36 tools one at a time, the A1 effect stays between 9.90 and 12.95 points on GPT-5-mini and between 11.19 and 14.40 points on Haiku~4.5. The cross-model comparison shows not much of a difference between GPT-5-mini and Haiku~4.5 on ASB.  The estimated cross-model difference in the A1 effect is $-1.54$ percentage points (95\% CI $[-12.45,+8.44]$), and for A3 it is $-1.70$ points ($[-12.00,+8.68]$). 

\begin{table}[t]
\centering
\caption{Robustness of the ASB A1 committed ASR effect to alternative
treatments of execution failures. Values are percentage-point changes
relative to A0 with 95\% cluster-bootstrap confidence intervals.}
\label{tab:asb_missingness}

\small
\setlength{\tabcolsep}{3pt}
\renewcommand{\arraystretch}{1.08}

\begin{tabularx}{\columnwidth}{@{}Xcc@{}}
\toprule
\textbf{Failure treatment} &
\textbf{GPT-5-mini} &
\textbf{Haiku~4.5} \\
\midrule

Drop failed runs
&
\shortstack{$+11.67$\\{\scriptsize $[+4.44,+19.07]$}}
&
\shortstack{$+13.21$\\{\scriptsize $[+6.33,+21.10]$}}
\\

Failure = not committed
&
\shortstack{$+11.67$\\{\scriptsize $[+4.44,+19.07]$}}
&
\shortstack{$+12.04$\\{\scriptsize $[+5.74,+19.07]$}}
\\

Failure = committed
&
\shortstack{$+11.67$\\{\scriptsize $[+4.44,+19.07]$}}
&
\shortstack{$+13.89$\\{\scriptsize $[+6.85,+21.85]$}}
\\

\bottomrule
\end{tabularx}

\vspace{2pt}
\begin{minipage}{0.98\columnwidth}
\scriptsize
GPT-5-mini recorded no execution failures, so all three treatments use
identical data and produce identical estimates and intervals.
\end{minipage}

\end{table}
\paragraph{The ASB result establishes sensitivity, but not its mechanism} From the results in Table~\ref{tab:asb_metrics}, we have established that changing how a threat is represented changes ASB's measured attack success rate. But we do not know exactly which part of the representation causes that change. The A1 names are less threat-related than A0, but they also differ in their token use. We noticed that on average, A1 uses fewer \texttt{o200k\_base} tokens than the original A0 names (average of 2.83 versus 3.47). The difference is even smaller with Haiku~4.5's tokenizer (average of 5.39 vs 5.86). Because both wording and form changed, we cannot say that threat-related wording alone caused the ASB effect. 

We also designed a more tightly matched neutral-name control intended to remove threat-related wording while remaining close to the original identifiers in token count, component structure, casing, and length. 
Out of 32 tools, 30 tools met the requirement for such a change, but only 10 of those passed the blind check for whether the new name still described the same tool function. Thus we did not run that arm. 

Thus, while we conclude that changing the threat representation changes ASB's attack success rate, we do not know exactly what part of the change causes that effect. 

\paragraph{Diagnostic A2 arm}
Since A2 alters the task specification and thus fails to meet our criteria of preserving threats, we used it for diagnostic purposes. On GPT-5-mini, A2 increases the committed ASR by $+4.44$ percentage points (95\% confidence interval $[+0.37,+9.44]$) and native ASR by $+2.78$ points ($[-0.56,+6.85]$, with no execution failures). For the Haiku~4.5 model, the committed ASR shift is $-2.23$ points ($[-11.17,+6.65]$) and native ASR shift is $-0.06$ points ($[-9.01,+8.68]$), but A2 has a 22.4\% execution failure rate (95\% CI $[15.0,29.8]$).

\subsection{MCPTox: Adding Threat-Related Tool Names}
Unlike ASB, where we removed threat-related wording, we change neutral names such as \texttt{qubit}, \texttt{stone}, and \texttt{fish} to more threat-related wording. We use 203 cases out of 1,348 cases for which we could reliably reproduce the benchmark's scoring rule. 

\paragraph{Adding threat-related wording lowers the attack success rate on GPT-5-mini} When we replaced the original neutral names with explicit threat-related names (arm N1), we saw a drop of about 11 percentage points (from 32.02\% to 21.02\%) in the attack success rate in GPT-5-mini. When we compare that to the more conservative change in threat wording of C1, it comes down by 2.96 points.

Haiku~4.5 shows a smaller shift, lowering the attack success rate by 4.11 percentage points (from 9.03\% to 4.92\%) for N1 vs N0 but has a similar effect with the more conservative change in threat wording of C1 with 2.63 points. Thus, the large shift appears most clearly for the explicit N1 naming intervention on GPT-5-mini.

\paragraph{Threat-related wording alone cannot explain the observed N1 effect}
The change of neutral names from N0 to threat-explicit wording in N1 not only changes the name and meaning but also the form. Compared to N0, N1 names are longer, contain more components, and use different casing. After observing initial N1 results, we added S3, a threat-neutral arm designed to match N1 on these measured properties without using threat-related words. Interestingly, we found that S3 alone lowers GPT-5-mini attack success rate by 8.54 percentage points.
N1 compared to S3 only lowers the attack success rate by 2.46 points. By contrast, N1 compared to the broader semantic-neutral controls S1/S2 lowers the attack success rate by 8.37 points.

This shows that the explicit threat wording may not be necessary for most of the GPT-5-mini effect.

\paragraph{Haiku~4.5 shows a different pattern} For Haiku~4.5, S3 leaves ASR nearly unchanged ($-0.16$ points), and N1 lowers ASR by 3.94 points relative to S3 (95\% CI $[-7.22,-0.82]$), although this contrast does not survive Holm correction. Haiku~4.5's baseline ASR is only 9.03\%, which leaves little room for large absolute changes. The matched neutral control therefore reproduces most of the effect on GPT-5-mini, while the evidence for Haiku~4.5 is inconclusive.

\paragraph{Multiplicity}
Our Holm family contains five contrasts per model, corrected within each model: C1 vs. N0, C1 vs. S1/S2, N1 vs. S1/S2, S1/S2 vs. the orthographic arms (O1, O2, NULL2), and N1 vs. S3. The last was added together with S3, after the initial N1 results (Section III-G). The comparisons of N1 and S3 with N0 in Table VII are descriptive and not part of this family. Of the ten tests, one remains significant:
N1 versus the semantic-neutral names on GPT-5-mini
($\Delta=-8.37$ points, $p_{\mathrm{adj}}<10^{-4}$).
N1 versus S3 on Haiku~4.5 has $\Delta=-3.94$ points
(95\% CI $[-7.22,-0.82]$, unadjusted $p=0.014$) but does not survive
correction ($p_{\mathrm{adj}}=0.068$).
All contrasts involving C1 have $p_{\mathrm{adj}}\geq 0.30$.

\paragraph{MCPTox shows that representation sensitivity can lower the ASR}
We saw in ASB that removing threat-related names increases the ASR and in MCPTox adding threat-related wording lowers the attack success rate. Based on the results we can say that representation sensitivity therefore does not have a fixed direction nor is it completely model agnostic.
MCPTox's results show clear representation sensitivity, but the exact feature responsible for that sensitivity remains only partly identified.
\begin{table}[t]
\centering
\caption{MCPTox representation effects. Values are percentage-point changes
in ASR with 95\% cluster-bootstrap confidence intervals over cases.
N0 is the original benchmark. Negative values indicate lower attack success.}
\label{tab:mcptox_results}

\footnotesize
\setlength{\tabcolsep}{2.5pt}
\renewcommand{\arraystretch}{1.08}

\begin{tabularx}{\columnwidth}{@{}Xlcc@{}}
\toprule
\textbf{Arm} &
\textbf{Compared with} &
\textbf{GPT-5-mini} &
\textbf{Haiku~4.5} \\
\midrule

O1 (orthographic) &
N0 &
\shortstack{$-0.99$\\{\scriptsize $[-4.27,+2.30]$}} &
\shortstack{$+0.33$\\{\scriptsize $[-1.64,+2.30]$}} \\

NULL2 (orthographic)$^\dagger$ &
N0 &
\shortstack{$-3.12$\\{\scriptsize $[-6.40,+0.16]$}} &
\shortstack{$+1.02$\\{\scriptsize $[-0.66,+2.85]$}} \\

O2 (orthographic) &
N0 &
\shortstack{$+1.31$\\{\scriptsize $[-2.30,+4.93]$}} &
\shortstack{$-0.99$\\{\scriptsize $[-3.12,+0.99]$}} \\

\addlinespace[2pt]

S1 (neutral) &
N0 &
\shortstack{$-2.96$\\{\scriptsize $[-7.22,+1.31]$}} &
\shortstack{$-1.48$\\{\scriptsize $[-4.60,+1.48]$}} \\

S2 (neutral) &
N0 &
\shortstack{$-2.30$\\{\scriptsize $[-6.73,+1.97]$}} &
\shortstack{$-1.64$\\{\scriptsize $[-4.93,+1.31]$}} \\

\textbf{S3 (matched neutral)} &
N0 &
\shortstack{$\mathbf{-8.54}$\\{\scriptsize $[-12.48,-4.76]$}} &
\shortstack{$-0.16$\\{\scriptsize $[-3.94,+3.61]$}} \\

\addlinespace[2pt]

C1 (conservative threat) &
N0 &
\shortstack{$-2.96$\\{\scriptsize $[-7.22,+1.15]$}} &
\shortstack{$-2.63$\\{\scriptsize $[-5.91,+0.33]$}} \\

\textbf{N1 (explicit threat)} &
N0 &
\shortstack{$\mathbf{-11.00}$\\{\scriptsize $[-15.60,-6.57]$}} &
\shortstack{$\mathbf{-4.11}$\\{\scriptsize $[-7.55,-0.99]$}} \\

\midrule

\textbf{N1 (explicit threat)} &
S1/S2$^\ast$ &
\shortstack{$\mathbf{-8.37}$\\{\scriptsize $[-11.25,-5.67]$}} &
\shortstack{$-2.55$\\{\scriptsize $[-5.50,+0.25]$}} \\

\textbf{N1 (explicit threat)} &
S3 &
\shortstack{$-2.46$\\{\scriptsize $[-5.58,+0.66]$}} &
\shortstack{$\mathbf{-3.94}$\\{\scriptsize $[-7.22,-0.82]$}} \\

\bottomrule
\end{tabularx}

\vspace{2pt}
\begin{minipage}{0.98\columnwidth}
\scriptsize
$^\dagger$O1 and NULL2 are byte-identical across all 203 cases; their
different estimates provide an empirical reference for run-to-run variability
under identical inputs.
$^\ast$S1/S2 denotes the semantic-neutral control contrast.
S3 is threat-neutral and matched to N1 on the measured identifier properties.
Multiplicity corrections apply to hypothesis tests rather than the displayed
confidence intervals; details are given in the analysis procedure.
\end{minipage}

\end{table}

\subsection{AgentDojo: Small ASR Effect, Utility Shift}


\begin{table}[t]
\centering
\caption{AgentDojo effects on security and benign utility.
Values are percentage-point changes relative to D0 with 95\% cluster-bootstrap confidence intervals.
}
\label{tab:agentdojo_results}
\small
\setlength{\tabcolsep}{4pt}
\renewcommand{\arraystretch}{1.08}

\begin{tabularx}{\columnwidth}{@{}Xcc@{}}
\toprule
\textbf{Outcome / intervention} &
\textbf{Effect} &
\textbf{95\% CI} \\
\midrule

\multicolumn{3}{@{}l}{\textit{Security: $\Delta$ASR}} \\

Dnull: orthographic
&
$-0.88$
&
$[-3.02,+1.25]$
\\

\textbf{Dval: required tool}
&
$\mathbf{-0.50}$
&
$\mathbf{[-3.65,+2.51]}$
\\

Dirrelevant: unrelated tool
&
$+0.88$
&
$[-1.00,+2.76]$
\\

\addlinespace

\multicolumn{3}{@{}l}{\textit{Benign utility under Dval}} \\

Requires renamed tool
&
$-5.36$
&
$[-10.00,-0.71]$
\\

Other tasks
&
$+5.24$
&
$[+3.51,+7.03]$
\\

\textbf{Difference}
&
$\mathbf{-10.59}$
&
$\mathbf{[-15.87,-5.69]}$
\\

\bottomrule
\end{tabularx}
\end{table}
AgentDojo differs from MCPTox and ASB in terms of how the attack reaches the agent. Unlike the other two benchmarks, AgentDojo does not have any malicious tool; instead, malicious instructions are injected into untrusted tool output to misuse the benign tools that are also needed for benign tasks. Thus, we add threat wording to a benign tool. For our experiment we only used the injection tasks whose attack path requires the renamed tool.

\paragraph{The observed ASR effect is small}
Adding threat wording to tool names produces only a small change in ASR. When we changed the names of the tools required by the attack, it caused the attack success rate to be lowered by 0.5 points going from 22.74\% to 22.24\% (95\% CI $[-3.65,+2.51]$). When we changed the name of a tool not required by the attack, its attack success rate increased by 0.88 points ($[-1.00,+2.76]$). The intervals are much smaller than the ASB and MCPTox results, thus providing a bounded contrast: large representational sensitivity is not inevitable for all benchmarks.

\paragraph{The same change affects benign utility}
Unlike the attack success rate, the benign utility showed a higher variation. Under Dval, it fell by 5.36 points (from 33.93\% to 28.57\%; 95\% CI $[-10.00,-0.71]$) on tasks that require the renamed tool, but rose by 5.24 points ($[+3.51,+7.03]$) on tasks that do not use it.
Because nothing those tasks depend on was changed, the rise indicates variation of similar magnitude that our design does not explain, such as temporal drift in the model API. We therefore treat the utility effect as suggestive rather than established.

One possible explanation for this behavior could be that the model just avoids the use of the renamed tool, but we observed that the renamed tool was being invoked in 99.6\% of runs under both Dval and D0 for relevant tasks. Thus, the change in benign utility is not explained by how often the renamed tool is invoked. 


While AgentDojo's result did not produce a large shift in attack success rate like ASB or MCPTox, it did show a possible shift in the benign utility. Representation sensitivity may therefore depend on what is being evaluated, not only on the benchmark.

\subsection{Cross-Benchmark Findings}
Across the three benchmarks on different models we saw that the representation sensitivity can be large but is not uniform. In ASB we saw removing threat wording makes the attack success rate jump; in MCPTox we saw that adding threat wording lowered the attack success rate, whereas in AgentDojo we saw smaller effect on attack success rate.
However, we do not interpret these differences as causal effects of benchmark architecture, since the benchmarks also differ in their attacks, models, and transformations.

Our MCPTox results also show that explicit threat words are not necessary for most of the GPT-5-mini effect.

Finally, we show that the representation sensitivity can depend on what the benchmark chooses to measure. In ASB, we saw the difference in committed ASR and native ASR; in AgentDojo, we saw only a small change in ASR and a larger change in the benign utility. 

Thus overall we can say that an agent's security score can depend not only on the underlying threat and model, but also on how that threat is represented and on how success is defined.

\section{Discussion}
\paragraph{TPRS}
TPRS tests if a benchmark's security score like ASR remains stable when the underlying security problem is held fixed but its agent-visible representation changes. We found that in ASB and MCPTox it was not stable and was comparatively stable on AgentDojo. Thus, security scores like ASR can be affected by the design decisions regarding representation.

The security conclusions obtained from one representation are not necessarily the same under another threat-preserving representation of the same security problem. With TPRS we propose a way to test such dependence before treating a benchmark's security score as evidence of the model's or attack's robustness. 
\paragraph{Implication}
Our work shows that a benchmark should
treat its representation choices as part of the evaluation design. If a benchmark is used to make a generalized security claim, we recommend that it evaluate with at least one threat-preserving
transformation and report any sensitivity as part
of the security metric. Similarly if a particular representation also affects the benign behavior, utility should also be measured with various threat-preserving transformations. Our AgentDojo results show why a security-only comparison can miss capturing the utility effect of representation. Benchmarks should also state explicitly when the attack is considered a success. Our ASB results show that native and
committed ASR capture different successes and do not exhibit identical sensitivity. In general, reporting both the benchmark configuration and its sensitivity to controlled representation changes makes clear which conclusions are properties of the underlying security problem and which may depend on how that problem is presented to the agent.
\paragraph{Interpretation of TPRS}
TPRS measures how much the measurement score like ASR changes under controlled representation changes, but its security interpretation can depend on who chooses that representation. When an attacker can choose the representation by picking the name of the tool or influencing the metadata associated with it (like in ASB or MCPTox), the representation can become part of the attack itself. In ASB, for example, the attacker supplies the malicious tool and chooses its name. A name such as \texttt{DNSConfiguration} may make the tool look benign, so A1 does not separate threat wording from apparent function (Section VI), but a neutral name costs the attacker nothing. Under this reading, ASB's original names understate the ASR an attacker can achieve by 11.67--13.21 points (committed ASR).
In such cases one should report TPRS as well as the maximum ASR observed when testing with different representations.  When a benchmark designer chooses the representation, TPRS characterizes how much the measurement score like ASR is dependent on benchmark design choices. The small change in ASR with AgentDojo  shows that the security measurement was comparatively stable under the tested representation changes.   

\paragraph{AgentDojo Result}
Even though AgentDojo's change in ASR was not significant, it provides a useful contrast to ASB and MCPTox. Its result shows that the large shift in sensitivity is not inevitable. Some benchmark measurements may appear much more sensitive to the representation than others. While we did not establish why AgentDojo's change was so small compared to the other two benchmarks, it does differ from ASB and MCPTox in several ways, including how the attack reaches the agent and how the harmful action is carried out.
One possibility could be that the representation matters more when the benchmark explicitly states the adversarial capability. We leave that testing to future work.

\section{Limitations and Future Work}
\paragraph{The tested transformations are not exhaustive}
We applied transformations on tool names and descriptions as they are among the most visible representations. But they are only one subset of all possible representations. There can be other types of transformation like argument names, tool schemas, instruction framings and so on. Thus we do not interpret our results as an interval of uncertainty over all the ways the same threat could be expressed.
Our future work may include testing and changing other aspects of threat representations while keeping the same security criteria. One can also try to develop transformations that better preserve the tool's described function while varying threat-related representation.

There is also a tradeoff between making the names more noticeably threat-related and maintaining the meaning of the representation. In our ASB experiment's A1 arm, the changes in the names were not independently validated for semantic equivalence. Thus, transformations may therefore not only alter the appearance of the threat but also the apparent functionality of the capability. When we tried to run a control experiment, this limitation became concrete. When designing a semantically equivalent arm, too few name changes were considered semantically equivalent, so that arm was not run. That is one of the reasons that the experiments were not able to clearly identify threat vocabulary as the main driver of the sensitivity.

\paragraph{Sensitivity attribution remains incomplete}
Our experiment with MCPTox does provide a stronger sensitivity control in the S3 arm where we match the threat-neutral names to the threat-related names of the N1 arm on identifier properties. For GPT-5-mini,
the S3 arm replicates most of N1's ASR shift, showing that the threat vocabulary is not necessary for most of that effect. The S3 arm does not isolate every property of an identifier; moreover, the tokenizer used for token count on GPT-5-mini may not be exact under Haiku~4.5's tokenizer. Thus the sensitivity attribution is not completely resolved. 

For future work one can design tighter controls that can vary semantics, tokenization, and structure of the identifier to determine to which properties of the representation the sensitivity can be attributed. 

\paragraph{The empirical scope is limited}
We evaluated three benchmarks and three model configurations: GPT-5-mini and Claude Haiku~4.5 on ASB and MCPTox, and GPT-4o-mini on AgentDojo. The experiments show the sensitivity to threat representation in the tested settings but do not establish its scale or generalization beyond the tested scope. 
Moreover, our 28,904 runs are not necessarily independent as the transformations applied are shared across runs.

We plan to extend this work across a larger set of models, benchmarks, attacks and also evaluate defenses. 

\paragraph{Ranking stability}
Since the benchmark scores are often used for comparisons, one important direction for future work could be to establish if model rankings or defense rankings reverse under representation changes. Testing that question would require multiple models or defenses with sufficient headroom under the same transformation family.
\section{Conclusion}
Agent-security benchmarks are often interpreted as measuring how robust a model is to an attack. Our results show that this measurement can depend on how the benchmark presents that attack to the LLM agent.
We introduced threat-preserving representation sensitivity (TPRS) as a measure of the effect of a
change in the representation of the attack seen by the agent while the underlying task, harmful action, security policy, ground truth, and evaluation criteria stay the same.

Across 28,904 agent runs on ASB, MCPTox, and AgentDojo, we found that TPRS can be large, can both increase and decrease security scores, and is not uniform across the different benchmarks. When we changed threat-related names to more neutral-sounding names, we saw an increase in committed ASR in ASB, whereas introducing threat-related names reduced ASR in MCPTox. The ASR effect observed in AgentDojo for a similar change in representation was much smaller, despite a benign utility change. On MCPTox, our controls showed that explicit threat vocabulary is not necessary for most of the GPT-5-mini effect.

We conclude that a benchmark measure of security may be dependent on representation decisions that are usually made implicitly. For any security claims intended to extend beyond one specific interface, benchmarks should evaluate the same under controlled threat-preserving variants rather than treating a single representation as measurement-neutral.

\bibliographystyle{IEEEtran}
\bibliography{reference}
\end{document}